\documentclass[onecollarge,fleqn,runningheads, preprint,showpacs,preprintnumbers,amsmath,amssymb]{svjour2}
\smartqed  
\usepackage{graphicx}
\usepackage{dcolumn}
\usepackage{bm}
\begin{document}

\title{Self-Similarity in Fully Developed Homogeneous Isotropic Turbulence Using the Lyapunov Analysis
}


\author{Nicola de Divitiis
}


\institute{Department of Mechanics and Aeronautics\\
University "La Sapienza", Rome, Italy \at
           via Eudossiana, 18  \\
              Tel.: +39-06-44585268\\
              Fax: +39-06-4881759\\
              \email{dedivitiis@dma.dma.uniroma1.it}           
}

\date{Received: date / Accepted: date}

\maketitle

\newcommand{\no}{\noindent}
\newcommand{\be}{\begin{equation}}
\newcommand{\ee}{\end{equation}}
\newcommand{\bea}{\begin{eqnarray}}
\newcommand{\eea}{\end{eqnarray}}
\newcommand{\bc}{\begin{center}}
\newcommand{\ec}{\end{center}}

\newcommand{\calr}{{\cal R}}
\newcommand{\calv}{{\cal V}}

\newcommand{\bff}{\mbox{\boldmath $f$}}
\newcommand{\bfg}{\mbox{\boldmath $g$}}
\newcommand{\bfh}{\mbox{\boldmath $h$}}
\newcommand{\bfi}{\mbox{\boldmath $i$}}
\newcommand{\bfm}{\mbox{\boldmath $m$}}
\newcommand{\bfp}{\mbox{\boldmath $p$}}
\newcommand{\bfr}{\mbox{\boldmath $r$}}
\newcommand{\bfu}{\mbox{\boldmath $u$}}
\newcommand{\bfv}{\mbox{\boldmath $v$}}
\newcommand{\bfx}{\mbox{\boldmath $x$}}
\newcommand{\bfy}{\mbox{\boldmath $y$}}
\newcommand{\bfw}{\mbox{\boldmath $w$}}
\newcommand{\bfk}{\mbox{\boldmath $\kappa$}}

\newcommand{\bfA}{\mbox{\boldmath $A$}}
\newcommand{\bfD}{\mbox{\boldmath $D$}}
\newcommand{\bfI}{\mbox{\boldmath $I$}}
\newcommand{\bfL}{\mbox{\boldmath $L$}}
\newcommand{\bfM}{\mbox{\boldmath $M$}}
\newcommand{\bfS}{\mbox{\boldmath $S$}}
\newcommand{\bfT}{\mbox{\boldmath $T$}}
\newcommand{\bfU}{\mbox{\boldmath $U$}}
\newcommand{\bfX}{\mbox{\boldmath $X$}}
\newcommand{\bfY}{\mbox{\boldmath $Y$}}
\newcommand{\bfK}{\mbox{\boldmath $K$}}

\newcommand{\bfrho}{\mbox{\boldmath $\rho$}}
\newcommand{\bfchi}{\mbox{\boldmath $\chi$}}
\newcommand{\bfphi}{\mbox{\boldmath $\phi$}}
\newcommand{\bfPhi}{\mbox{\boldmath $\Phi$}}
\newcommand{\bflambda}{\mbox{\boldmath $\lambda$}}
\newcommand{\bfxi}{\mbox{\boldmath $\xi$}}
\newcommand{\bfLambda}{\mbox{\boldmath $\Lambda$}}
\newcommand{\bfPsi}{\mbox{\boldmath $\Psi$}}
\newcommand{\bfomega}{\mbox{\boldmath $\omega$}}
\newcommand{\bfeps}{\mbox{\boldmath $\varepsilon$}}
\newcommand{\bfepsn}{\mbox{\boldmath $\epsilon$}}
\newcommand{\bfzeta}{\mbox{\boldmath $\zeta$}}
\newcommand{\bfkappa}{\mbox{\boldmath $\kappa$}}
\newcommand{\itPsi}{\mbox{\it $\Psi$}}
\newcommand{\itPhi}{\mbox{\it $\Phi$}}
\newcommand{\bint}{\mbox{ \int{a}{b}} }
\newcommand{\ds}{\displaystyle}
\newcommand{\Sum}{\Large \sum}

\begin{abstract}
In this work,
we calculate the self-similar longitudinal velocity correlation function and the statistical properties of velocity difference using the results of the Lyapunov analysis of the fully developed isotropic homogeneous turbulence just presented by the author in a previous work \cite{deDivitiis2009}. There, a closure of the von K\'arm\'an-Howarth equation is proposed and
the statistics of velocity difference is determined through a specific analysis of the Fourier-transformed Navier-Stokes equations.

The correlation functions correspond to steady-state solutions of the von K\'arm\'an-Howarth equation 
under the self-similarity hypothesis introduced by von K\'arm\'an.
These solutions are numerically determined with the statistics of velocity difference.
The obtained results adequately describe the
several properties of the fully developed isotropic turbulence.
\keywords{Self-Similarity \and Lyapunov Analysis \and von K\'arm\'an-Howarth equation \and Velocity difference statistics}
\PACS{47.27.-i}
\end{abstract}

\section{\bf Introduction}

A recent work of the author dealing with the homogeneous isotropic turbulence \cite{deDivitiis2009}, suggests a novel method to analyze the fully developed turbulence through a specific Lyapunov analysis of the relative motion of two fluid particles.
The analysis expresses the velocity fluctuation as the combined effect of the exponential growth rate of the fluid velocity in the Lyapunov basis, and of the rotation of the same basis with respect to the fixed frame of reference. 
The results of this analysis lead to the closure of the von K\'arm\'an-Howarth equation and give an explanation of the mechanism of the energy cascade. A constant skewness of the velocity derivative 
$\partial u_r/ \partial r$ is calculated which is in agreement with the various source of the literature.
Moreover, the statistics of the velocity difference can be inferred looking at the Fourier series of the velocity.
This is a non-Gaussian statistics, where the constancy of the skewness of 
$\partial u_r/ \partial r$ implies that the other higher absolute moments increase with the Taylor-scale Reynolds number. 

The present work represents a further contribution of Ref. \cite{deDivitiis2009}.
Here, the self-similar solutions of the von K\'arm\'an-Howarth equation are numerically calculated 
using the closure obtained in the previous work and the statistics of the velocity difference is obtained.

\section{\bf Analysis \label{s1}}

For sake of convenience, this section reports the main results of the Lyapunov
analysis obtained in the Ref. \cite{deDivitiis2009}.

\bigskip

As well known, the pair correlation function $f$ of the longitudinal velocity $u_r$
for the fully developed isotropic and homogeneous turbulence, satisfies the von K\'arm\'an-Howarth equation \cite{Karman38}
\bea
\ds \frac{\partial f}{\partial t} = 
\ds  \frac{K(r)}{u^2} +
\ds 2 \nu  \left(  \frac{\partial^2 f} {\partial r^2} +
\ds \frac{4}{r} \frac{\partial f}{\partial r}  \right) -10 \nu f \frac{\partial^2 f}{\partial r^2}(0)
\label{vk-h}  
\eea
the boundary conditions of which are
\bea
\begin{array}{l@{\hspace{+0.2cm}}l}
\ds f(0) = 1, \ \frac{\partial f (0)} {\partial r} = 0 \\\\
\ds \lim_{r \rightarrow \infty} f (r) = 0
\end{array}
\label{bc0}
\eea
Into Eq. (\ref{vk-h}), $r$ is the separation distance and $u$ is the standard deviation of $u_r$, which satisfies \cite{Karman38}, \cite{Batchelor53}
\bea
\ds \frac{d u^2}{d t} = 10 \nu u^2 \frac{\partial^2 f}{\partial r^2}(0) 
\eea
This equation gives the rate of the kinetic energy and is determined putting $r=0$ in the von K\'arm\'an-Howarth equation \cite{Karman38}, \cite{Batchelor53}.
The function $K(r)$,  related to the triple velocity correlation function, 
represents the effect of the inertia forces and expresses the mechanism of energy cascade.
In accordance to the Lyapunov analysis presented in Ref \cite{deDivitiis2009}, 
the analytical expression of $K(r)$ is in terms of $f$ and on its space derivative 
\bea
\ds K(r) = u^3 \sqrt{\frac{1-f}{2}} \frac{\partial f}{\partial r}
\label{K}
\eea
The proposed expression of $K(r)$ satisfies the two conditions
$\partial K/\partial r (0) = 0$, $K(0) = 0$,
which represent, respectively the condition of homogeneity and 
the fact that $K$ does not modify the fluid kinetic energy \cite{Karman38}, \cite{Batchelor53}.
The skewness of the longitudinal velocity difference $\Delta u_r$ is calculated as  \cite{Batchelor53}
\bea
\ds H_3(r) = \frac{\left\langle ( \Delta u_r )^3 \right\rangle} 
{\left\langle (\Delta u_r)^2\right\rangle^{3/2}} =
  \frac{6 k(r)}{\left( 2 (1 -f(r)  )   \right)^{3/2} }
\label{H_3_01}
\eea 
where $k(r)$ is the longitudinal triple velocity correlation function, 
related to $K(r)$ through \cite{Batchelor53}
\bea
K(r)= u^3 \left(  \frac{\partial}{\partial r}  + \frac{4}{r}  \right) k(r)
\label{kk}
\eea
The result of this analysis is that the skewness of $\partial u_r/\partial r$ is a constant which does not depend on the Reynolds number, whose value is $H_3(0) = -3/7$ \cite{deDivitiis2009}.
The solutions of the von K\'arm\'an-Howarth equation provide second and third dimensionless statistical moments of $\Delta u_r$. In line with the analysis of Ref.\cite{deDivitiis2009}, 
the higher moments are consequentely determined, taking into account that the analytical structure of $\Delta u_r$ is  expressed as 
\bea
\begin{array}{l@{\hspace{+0.2cm}}l}
\ds \frac {\Delta {u}_r}{\sqrt{\langle (\Delta {u}_r)^2} \rangle} =
\ds \frac{   {\xi} + \psi \left( \chi ( {\eta}^2-1 )  -  
\ds  ( {\zeta}^2-1 )  \right) }
{\sqrt{1+2  \psi^2 \left( 1+ \chi^2 \right)} } 
\end{array}
\label{fluc4}
\eea 
Equation (\ref{fluc4}), which arises from statistical considerations about the 
Fourier-transformed Navier-Stokes equations in fully developed turbulence, 
expresses the internal structure of the isotropic turbulence, 
where  $\xi$, ${\eta}$ and $\zeta$ are independent  centered random variables which exhibit the gaussian distribution functions $p(\xi)$, $p(\eta)$ and $p(\zeta)$ whose  standard deviation is equal to the unity.
Thus, the moments of $\Delta {u}_r$ are easily calculated from Eq. (\ref{fluc4}) \cite{deDivitiis2009}
\bea
\begin{array}{l@{\hspace{+0.2cm}}l}
\ds H_n \equiv \frac{\left\langle (\Delta u_r)^n \right\rangle}
{\left\langle (\Delta  u_r)^2\right\rangle^{n/2} }
= 
\ds \frac{1} {(1+2  \psi^2 \left( 1+ \chi^2 \right))^{n/2}} 
\ds \sum_{k=0}^n 
\left(\begin{array}{c}
n  \\
k
\end{array}\right)  \psi^k
 \langle \xi^{n-k} \rangle 
  \langle (\chi(\eta^2 -1) - (\zeta^2 -1 ) )^k \rangle 
\end{array}
\label{m1}
\eea
where
\bea
\begin{array}{l@{\hspace{+0.2cm}}l}
\ds   \langle (\chi(\eta^2 -1) - (\zeta^2 -1 ) )^k \rangle = 
\ds \sum_{i=0}^k 
\left(\begin{array}{c}
k  \\
i
\end{array}\right)  
(-\chi)^i 
 \langle (\zeta^2 -1 )^i \rangle 
 \langle (\eta^2 -1 )^{k-i} \rangle \\\\
\ds  \langle (\eta^2 -1 )^{i} \rangle = 
\sum_{l=0}^i 
\left(\begin{array}{c}
i  \\
l
\end{array}\right)  
(-1)^{l}
\langle \eta^{2(i-l)} \rangle 
 \end{array}
\label{m2}
\eea
In particular, the third moment or skewness, $H_3$,
which is related for the energy cascade, is
\bea
\ds H_3= \frac{  8  \psi^3 \left( \chi^3 - 1 \right) }
 {\left( 1+2  \psi^2 \left( 1+ \chi^2 \right) \right)^{3/2}  }
\label{H_3}
\eea 
As far as $\psi$ is concerned, this is a function of the separation distance and of the Reynolds number \cite{deDivitiis2009}
\bea
\psi({\bf r}, R) =  
\sqrt{\frac{R}{15 \sqrt{15}}} \
\hat{\psi}(r)
\label{Rl}
\eea
where $R = u \lambda_T / \nu$ and $\ds \lambda_T =u/\sqrt{\langle (\partial u_r / \partial r)^2 \rangle}$
are, respectively, the Taylor-scale Reynolds number and the Taylor scale, whereas the 
function $\hat{\psi}(r)$ is determined through Eq. (\ref{H_3}) as soon as $H_3(r)$ is known. 
The parameter $\chi$ is also a function of $R$ which is implicitly calculated 
putting $r=0$ into Eq. (\ref{H_3}), i.e. \cite{deDivitiis2009}
\bea
\frac{  8  {\psi_0}^3 \left(  1-\chi^3 \right) }
 {\left( 1+2  {\psi_0}^2 \left( 1+ \chi^2 \right) \right)^{3/2}  }
= \frac{3}{7}
\label{sk1}
\eea
with ${\psi_0} = \psi(R,0)$ and $\hat{\psi_0} = 1.075$ \cite{deDivitiis2009}.
\\
From Eqs. (\ref{fluc4}) and (\ref{Rl}), all the absolute values of the dimensionless 
moments of $\Delta u_r$ of order greater than $3$ rise with $R$, indicating that the intermittency increases with the Reynolds number.
\\
The PDF of $\Delta u_r$ can be formally expressed through distribution functions $p(\xi)$, $p(\eta)$ and $p(\zeta)$, using the Frobenious-Perron equation
\bea
\begin{array}{l@{\hspace{+0.3cm}}l}
F(\Delta {u'}_r) = 
\ds \int_\xi 
\int_\eta  
\int_\zeta 
p(\xi) p(\eta) p(\zeta) \
\delta \left( \Delta u_r-\Delta {u'}_r \right)   
d \xi d \eta d \zeta
\end{array}
\label{frobenious_perron}
\eea 
where $\delta$ is the Dirac delta.

The spectrums $E(\kappa)$ and $T(\kappa)$ are the Fourier Transforms of $f$ and $K$ \cite{Batchelor53}, respectively
\bea
\left[\begin{array}{c}
\ds E(\kappa) \\\\
\ds T(\kappa)
\end{array}\right]  
= 
 \frac{1}{\pi} 
 \int_0^{\infty} 
\left[\begin{array}{c}
 \ds  u^2 f(r) \\\\
 \ds K(r)
\end{array}\right]  \kappa^2 r^2 
\left( \frac{\sin \kappa r }{\kappa r} - \cos \kappa r  \right) d r 
\label{Ek}
\eea

\section{\bf Self-Similarity \label{s2}}

An ordinary differential equation for describing the spatial evolution of $f$ is derived
from the von K\'arm\'an-Howarth equation under the hypothesis of self-similarity
and using the closure given by Eq. (\ref{K}). This equation represents a boundary problem which is  
then transformed into an initial condition problem in the variable $r$.

\bigskip

Far from the initial condition, it is reasonable that the simultaneous effects of the
mechanism of the cascade of energy and of the viscosity act keeping $f$ and $E(\kappa)$  
similar in the time. This is the idea of self-preserving correlation function and turbulence 
spectrum which was originally introduced by von K\'arm\'an (see ref. \cite{Karman49} and reference therein).

In order to analyse this self-similarity, it is convenient to express $f$ in terms of the dimensionless variables 
$
\ds  \hat{r} = r / \lambda_T
$
and 
$ \ds  \hat{t}  = t u /  \lambda_T$, i.e., $f = f(\hat{t}, \hat{r})$.
As a result, Eq. (\ref{vk-h}) reads as follows
\bea
\ds \frac{\partial f}{\partial \hat{t}} \ \frac{\lambda_T}{u} \frac{d}{dt} 
\left( \frac{t u}{\lambda_T} \right) -
\ds \frac{\partial f}{\partial \hat{r}} \ \frac{\hat{r}}{u} \frac{d \lambda_T}{dt}  = 
\ds  \sqrt{\frac{1-f}{2}} \frac{\partial f}{\partial \hat{r}}+
\ds  \frac{2}{R}  \left(  \frac{\partial^2 f} {\partial \hat{r}^2} +
\ds \frac{4}{\hat{r}} \frac{\partial f}{\partial \hat{r}}  \right) + \frac{10}{R} f 
\label{vk-h_adim}  
\eea
where
$
{\partial^2 f} / {\partial \hat{r}^2}(0) \equiv -1
$.
This is a non--linear partial differential equation whose 
coefficients vary in time according to the rate of kinetic energy
\bea
\ds \frac{d u^2}{d t} = - \frac{10 \nu u^2}{\lambda_T^2}
\label{rate_energy}
\eea
If the self--similarity is assumed, 
all the coefficients of Eq. (\ref{vk-h_adim}) must not vary with the time \cite{Karman38}, \cite{Karman49}, thus one obtains
\bea
R = \mbox{const}
\label{coeff1}
\eea
\bea
 a_1 \equiv 
\frac{\lambda_T}{u} \frac{d}{dt} \left( \frac{t u}{\lambda_T} \right) = \mbox{const}
\label{coeff2}
\eea
\bea
a_2 \equiv \frac{1}{u} \frac{d \lambda_T}{dt} = \mbox{const}, \ \
\label{coeff3}
\eea
From  Eqs. (\ref{rate_energy}) and (\ref{coeff1}), 
$\lambda_T$ and $u$ will depend upon the time according to 
\bea
\begin{array}{l@{\hspace{-0.cm}}l}
\ds \lambda_T(t) = \lambda_T(0)\sqrt{1 +10 \nu /\lambda_T^2(0) t }, \ \ \ \ 
\ds u (t)= \frac{u(0)} {\sqrt{1+10 \nu /\lambda_T^2(0) \ t}}.
\end{array}
\label{ult}
\eea
The corresponding values of $a_1$ and $a_2$ are determined substituting these latter expressions
into Eqs. (\ref{coeff2}) and (\ref{coeff3}), respectively, i.e.
\bea
\begin{array}{l@{\hspace{-0.cm}}l}
\ds a_1 \equiv \frac{\lambda_T}{u} \frac{d}{dt} \left( \frac{t u}{\lambda_T} \right) =
\ds \frac{1}{1+10 \nu / \lambda_T^2(0) t} \\\\
\ds a_2 \equiv \frac{1}{u} \frac{d \lambda_T}{dt} = \frac{5}{R} 
\end{array}
\eea
The coefficient $a_1$ decreases with the time and for $t \rightarrow \infty$, $a_1 \rightarrow 0$,
whereas $a_2$ remains constant.
Therefore, for $t \rightarrow \infty$, one obtains the self-similar correlation function 
$f (\hat{r})$, which does not depend on the initial condition and that obeys to 
the following non--linear ordinary differential equation 
\bea
\begin{array}{l@{\hspace{-0.cm}}l}
\ds \frac{5}{R} \ \frac{d f} {d \hat{r}} \ \hat{r} +
\ds  \sqrt{\frac{1-f}{2}} \ \frac{d f} {d \hat{r}}  +
\ds \frac{2}{R}  \left(  \frac{d^2 f} {d \hat{r}^2} +
\ds \frac{4}{\hat{r}} \frac{d f}{d \hat{r}}  \right) + \frac{10}{R}  f = 0
\end{array}
\label{vk-h0}  
\eea
The first term of Eq. (\ref{vk-h0}) represents the time derivative of $f$
and does not influence the mechanism of energy cascade.
In line with von K\'arm\'an \cite{Karman38}, \cite{Karman49}, we search the self--similar 
solutions over the whole range of $\hat{r}$, with the exception of the dimensionless distances whose order magnitude exceed $R$.
This corresponds to assume the self--similarity for all the frequencies of the energy spectrum, 
but for the lowest ones \cite{Karman38}, \cite{Karman49}. 
Accordingly,  the self-similar solutions of the von K\'arm\'an-Howarth equation 
are also steady solutions, thus the first term of Eq. (\ref{vk-h0}) can be neglected with respect to the other ones, i.e.  
\bea
\begin{array}{l@{\hspace{-0.cm}}l}
\ds  \sqrt{\frac{1-f}{2}} \ \frac{d f} {d \hat{r}}  +
\ds \frac{2}{R}  \left(  \frac{d^2 f} {d \hat{r}^2} +
\ds \frac{4}{\hat{r}} \frac{d f}{d \hat{r}}  \right) + \frac{10}{R}  f = 0
\end{array}
\label{vk-h1}  
\eea
The boundary conditions of Eq. (\ref{vk-h1}) are from Eqs. (\ref{bc0}) \cite{Karman38}
\bea
f(0) =   1, \ \frac{d f(0)} {d \hat{r}} = 0
\label{bc1}
\eea
\bea
\lim_{\hat{r} \rightarrow \infty} f(\hat{r} ) = 0
\label{bc2}
\eea
Since the solutions $f \in C^2(0, \infty)$ 
 exponentially tend to zero when $r \rightarrow \infty$, and $\lambda_T$ is considered to be an assigned
 quantity, the boundary condition (\ref{bc2}) can be replaced by the following condition in the
  origin
\bea 
\ds \frac{d^2 f (0)} {d \hat{r}^2} = -1
\label{bc3}
\eea

Therefore, the boundary problem represented by Eqs. (\ref{vk-h1}),  (\ref{bc1}) and (\ref{bc2}), corresponds to the following initial condition problem written in the normal form
\bea
\begin{array}{l@{\hspace{+0.cm}}l}
\ds  \frac{d f}{d \hat{r}} =  F \\\\
\ds \frac{d F}{d\hat{r}} = -5 f -
\left( \frac{1}{2} \sqrt{\frac{1-f}{2}} R + \frac{4}{\hat{r}} \right) F
\end{array}
\label{vk-h2}  
\eea
the initial condition of which is 
\bea 
\ds f(0) = 1,  F(0) = 0
\label{ic}
\eea
This ordinary differential system arises from Eq. (\ref{vk-h1}), 
where according to Eq. (\ref{bc3}), one must take into account that
\bea
\ds \lim_{\hat{r} \rightarrow 0} \frac{F(\hat{r})}{\hat{r}} = \lim_{\hat{r} \rightarrow 0} \frac{dF(\hat{r})}{d\hat{r}} = -1
\eea

\section{\bf Results and Discussion \label{s3}}

In this section,
the solutions of the system (\ref{vk-h2}) are qualitatively studied and numerically calculated
for several values of the Reynolds numbers, and
the statistics of the velocity difference is investigated through the analysis seen in the section \ref{s1} (Eqs. (\ref{fluc4})-(\ref{frobenious_perron})).

\bigskip


The  analysis of Eqs. (\ref{vk-h2}) shows that, for large $\hat{r}$, $f$ exponentially decreases, 
thus all the integral scales of $f$ are finite quantities and the energy spectrum is a definite quantity whose integral over the Fourier space gives the turbulent kinetic energy.

It is worth to remark that, where $K(r)$ is about constant, $f$ behaves like 
\bea 
\ds f -1 \approx \hat{r}^{2/3}
\eea 
as a result, the energy spectrum varies according to the Kolmogorov law $\kappa^{-5/3}$ 
and the interval where this happens, identifies the inertial subrange of Kolmogorov.

Furthermore, the solutions $f \in C^6(0, \infty)$, of the system  (\ref{vk-h2})-(\ref{ic}),
in the vicinity of the origin read as 
\bea
\ds f(\hat{r}) = 1- \frac{\hat{r}^2}{2} + \frac{10+R}{112} \hat{r}^4 + O(\hat{r}^6)
\label{fapprox0}
\eea
The more rapid variations of $f$ and of its derivative occur in close proximity of the origin \cite{Batchelor53} and these are responsible for the behavior of $E(\kappa)$ at high wavenumbers.
Thus, we assume that the minimum length scale of $f$ is due to the second and the third term of Eq. (\ref{fapprox0}). Under this scale the effects of the inertia and pressure forces are
negligible with respect to the viscous forces.
This scale is determined as two times the separation distance $\hat{r}>0$ where the first derivative
of $\ds - \hat{r}^2/{2} + (10+R)/{112} \ \hat{r}^4$ vanishes.
Then, the corresponding wave-number (minimum scale) is related to $R$ through the relationship
\bea
\ds \kappa_{max} \approx \frac{2 \pi}{\lambda_T} \sqrt{\frac{10 +R}{112}}
\label{kmax}
\eea 
For higher wavenumbers, in the dissipative range, the viscosity influences $f$ in such a way that $E(\kappa)$ decreases much more rapidly with respect to the inertial subrange.  In the dissipative interval, $E(\kappa)$ roughly coincides with the Fourier transformation of  $f\simeq 1- \hat{r}^2/2$ $\simeq 1/ ( 1+ \hat{r}^2/2)$, and, for this reason, one could expect that the energy spectrum behaves like
\cite{Debnath}
\bea
E(\kappa) \approx \exp(-a \kappa)
\label{E7}
\eea
for $\kappa > \kappa_{max}$, where $a = O(1) >0$ is a proper parameter which depends upon the 
Reynolds number.
The value of $\kappa_{max}$, calculated following Eq. (\ref{kmax}), should indicate
the order of magnitude of the separation wavenumber between the Kolmogorov inertial 
subrange and the dissipative interval.

\bigskip

Now, several numerical solutions of Eqs. (\ref{vk-h2}) were calculated for different values of the Taylor-scale Reynolds numbers by means of the fourth-order Runge-Kutta scheme of integration.
\suppressfloats
 \begin{figure}[t]
	\centering
         \includegraphics[width=0.45\textwidth]{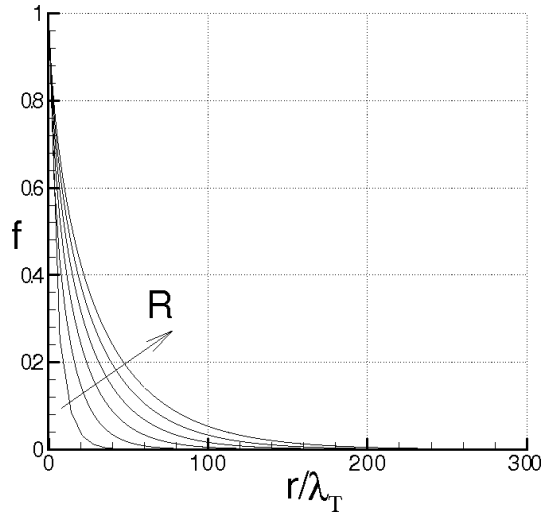}
\caption{Longitudinal correlation function for different Taylor-Scale Reynolds numbers.}
\label{figura_1}
\end{figure}
\begin{figure}[t]
	\centering
         \includegraphics[width=0.45\textwidth]{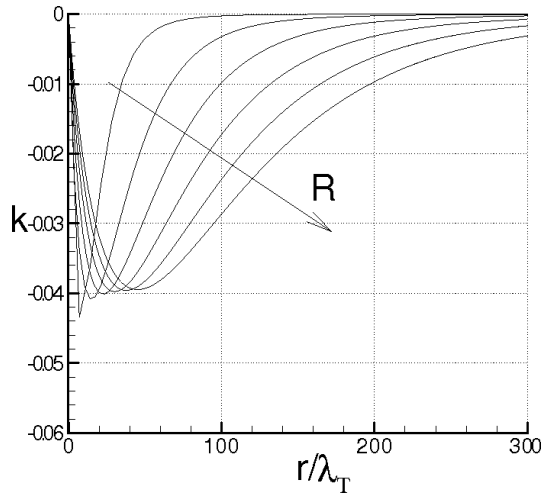}
\caption{Longitudinal triple correlation function for different Taylor-Scale Reynolds numbers.}
\label{figura_2}
\end{figure}

The cases here analyzed correspond to $R$ = $100$, $200$, $300$, $400$, $500$ and $600$.
The fixed step size of the integrator scheme is selected on the basis of the asymptotic
stability condition of Eq. (\ref{vk-h2}). This is $\Delta \hat{r} = \sqrt{2}/R$ \cite{NAG} and 
provides a fairly accurate description of the correlation function at small scales.

Figures \ref{figura_1} and \ref{figura_2} show the numerical solutions of Eqs. (\ref{vk-h2}), where double and triple longitudinal correlation functions are both in terms of $\hat{r}$, for the different values of $R$. Due to the mechanism of energy cascade,
represented by the expression of $K(r)$ (see Eq. (\ref{K})), the tail of $f$ rises with $R$ 
accordingly to Eq. (\ref{fapprox0}) and this means that, for an assigned value of 
 $\lambda_T$, all the integral scales of $f$ rise with the Reynolds number.
The triple correlation function is shown in terms of $\hat{r}$ in 
Fig. \ref{figura_2}. Because of energy cascade, and according to Eq. (\ref{K}),
 $k$ decays more slowly than $f$, for $\hat{r} \rightarrow \infty$. 
It is apparent that assigned variations of $k$ correspond to variations of $\hat{r}$  whose size increases with $R$.
The maximum of $\vert k \vert$ gives the entity of the mechanism of energy cascade.
This is slightly less than 0.05 and agrees quite well with the numerous data of the literature  
(see \cite{Batchelor53} and Refs. therein).

\begin{figure}[t]
	\centering
         \includegraphics[width=0.47\textwidth]{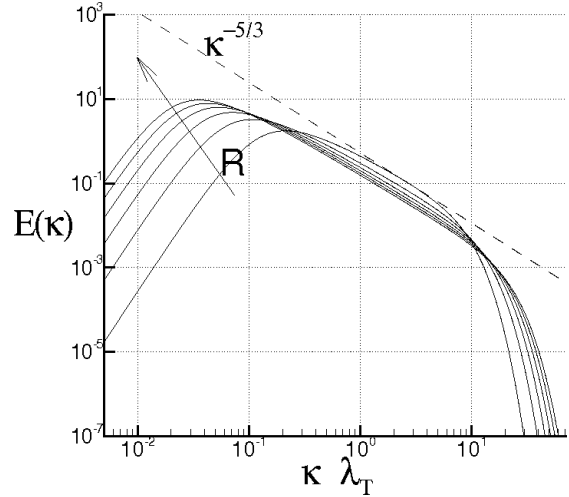}
\caption{Turbulent Energy Spectrum for different Taylor-Scale Reynolds numbers.}
\label{figura_3}
\end{figure}

\begin{figure}[t]
	\centering
         \includegraphics[width=0.47\textwidth]{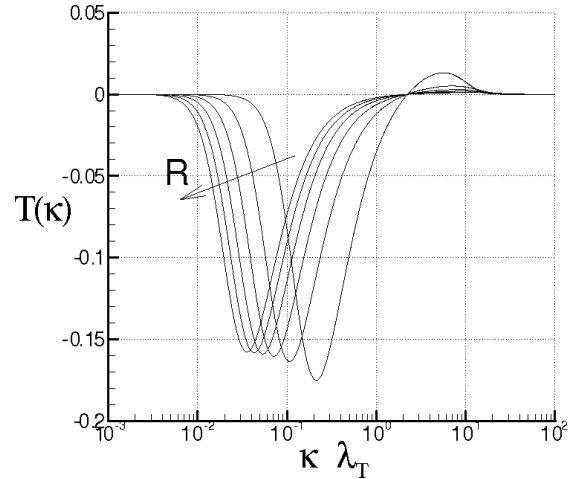}
\caption{"Transfer function $T(\kappa)$" for several Taylor-Scale Reynolds numbers.}
\label{figura_4}
\end{figure}

Figures \ref{figura_3} and \ref{figura_4} show the plots of $E(\kappa)$ and $T(\kappa)$ 
calculated with Eq. (\ref{Ek}), for the same Reynolds numbers.
As the consequence of the mathematical properties of $f$, the energy spectrum behaves like 
$E(\kappa) = O(\kappa^4)$ in proximity of the origin, and after a maximum is about parallel to the $-5/3$ Kolmogorov law (dashed line in Fig. \ref{figura_3}) in a given interval of the wave-numbers.
This interval defines the inertial range of Kolmogorov, and its size increases with $R$. 
This arises from the fact that there exists a spatial interval where $K(r)$ is about constant
and, as seen, this determines that $f -1 \approx r^{2/3}$.
At higher wave-numbers,  the energy spectrum decreases more rapidly with respect to the Kolmogorov subrange and its variations almost agree with Eq. (\ref{E7}).
The constant $a$ of Eq. (\ref{E7}) varies from about 0.53 to  0.26 
when $R$ assumes the values of 100 and 600, respectively.
The wavenumber of separation between the two regions changes with the Reynolds number
and follows the variations previously determined with Eq. (\ref{kmax}).
\suppressfloats
Since $K$ does not modify the kinetic energy of the flow, according to Eq. (\ref{K}), 
the integral of $T(\kappa)$ over $\kappa$ results to be identically equal to zero
at all the Reynolds numbers.

\begin{figure}[t]
	\centering
         \includegraphics[width=0.45\textwidth]{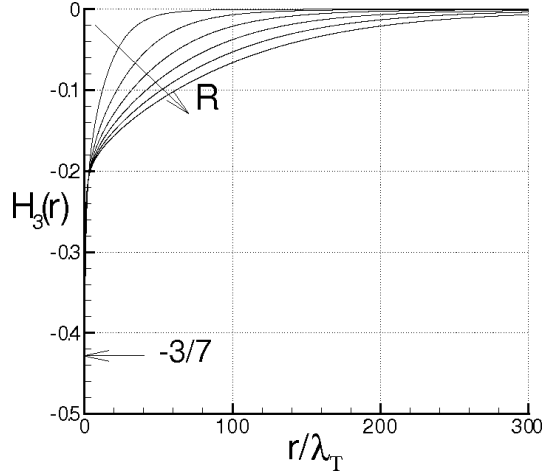}
\caption{Skewness of $\Delta u_r$ at different Taylor-Scale Reynolds numbers.}
\label{figura_5}
\end{figure}

\begin{figure}[t]
	\centering
         \includegraphics[width=0.45\textwidth]{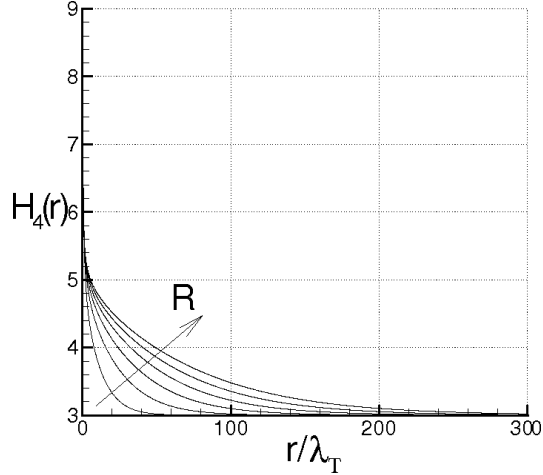}
\caption{Flatness of $\Delta u_r$ at different Taylor-Scale Reynolds numbers.}
\label{figura_6}
\end{figure}

In the Figs. \ref{figura_5} and \ref{figura_6}, skewness and flatness of 
$\Delta u_r$ are shown in terms of $\hat{r}$ for the same values of the Reynolds numbers.
The skewness $H_3$ is first calculated according to Eq. (\ref{H_3_01}) and 
thereafter the flatness $H_4$ has been determined using Eq. (\ref{m1}). 
For a given value of $R$, $\vert H_3 \vert$ starts from 3/7 at the origin, then decreases to small values, while $H_4$ starts from values quite greater than 3 at $r=0$, then reaches the value of 3. In line with Eq. (\ref{K}), since $f$ decays more rapidly than $k$  as $r \rightarrow \infty$, $H_4$ goes to 3 more fastly than $H_3$ tending to zero.
Although $H_3(0)$ does not depend upon $R$, $H_3(\hat{r})$ is a rising function
of the Reynolds number and, in any case, the intermittency of $\Delta u_r$ increases with $R$
according to Eqs. (\ref{fluc4})-(\ref{Rl}).

\begin{figure}[t]
	\centering
         \includegraphics[width=0.45\textwidth]{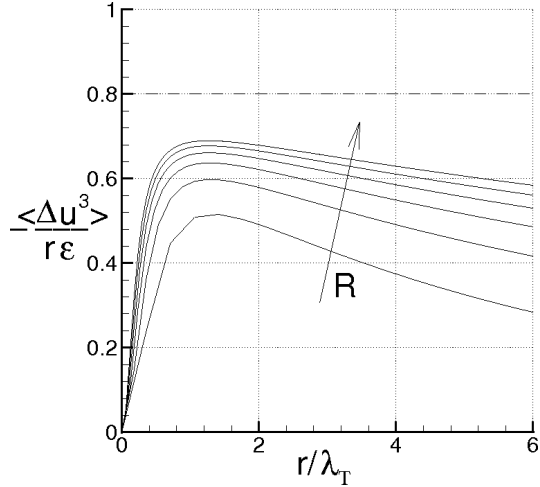}
\caption{Kolmogorov function for several Taylor-Scale Reynolds numbers.}
\label{figura_7}
\end{figure}

\begin{figure}[t]
     \centering
     \includegraphics[width=0.55\textwidth]{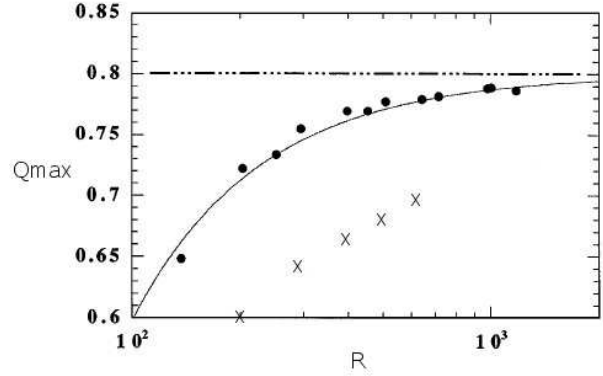}
\caption{Maximum of the Kolmogorov Function in terms of the Taylor-Scale Reynolds number.
These data are from Ref. \cite{Moisy99}. The symbols "X" are for the present result.}
\label{figura_7a}
\end{figure}

Next, the Kolmogorov function $Q(r)$ and the Kolmogorov constant $C$, are determined using 
the previous results. 
According to the theory, the Kolmogorov function, defined as
\bea 
\ds Q(r) = - \frac{\langle (\Delta u_r)^3 \rangle} { r \varepsilon}
\label{k_f}
\eea
is constant with respect to $r$, and is equal to $4/5$ as long as $r/\lambda_T = O(1)$, where
$\ds \varepsilon = - 3/2 \ d u^2/ dt$ is the rate of the energy of dissipation.
In Fig. \ref{figura_7}, $Q(r)$ is calculated through the skewness of velocity difference 
and is shown in terms of $\hat{r}$. This exhibits a maximum for $\hat{r} = O(1)$ and
quite small variations for higher $\hat{r}$, as the Reynolds number increases. 
These behavior is the consequence of the aforementioned variations of $H_3(r)$ with $R$
and $r$. 
The maximum of $Q$ rises with $R$ and seems to tend toward the limit $4/5$ 
prescribed by the Kolmogorov theory. 
In Fig. \ref{figura_7a}, the maximum of the Kolmogorov function  is in term of $R$.
These data  ("x" symbols) are shown in comparison with those 
of Ref. \cite{Moisy99} (circular filled symbols and continuous line). 
\suppressfloats
There, the Kolmogorov equation with a forcing term, is compared with experimental measurements, 
for helium gas at low temperature. 
It is apparent that, for assigned $R$, the corresponding values of $Q_{max}$, calculated with 
Eq. (\ref{k_f}), are less than those of Ref. \cite{Moisy99}, and such difference varies with $R$ with an average percentage value of about $12 \%$. 
This disagreement could be due to the self-similarity here assumed or to possible differences in the estimation of the Taylor-scale. 
Specifically, $\lambda_T$ is here analytically calculated or assumed, whereas in Ref. \cite{Moisy99}, it is determined through measurement of dissipation rate.
Nevertheless, the two set of data can be considered comparable, since the two variations 
almost exhibit the same trend.
\suppressfloats
\suppressfloats
\begin{table}[t]
\centering
\caption{Kolmogorov constant for different Taylor-Scale Reynolds number.}
  \begin{tabular}{cc} 
$R$ \       &  \ $C$ \\[2pt] 
\hline
\hline
100 \       & \ 1.8860      \\
200 \       & \ 1.9451      \\
300 \       & \ 1.9704      \\
400 \       & \ 1.9847      \\
500 \       & \ 1.9940    \\
600 \       & \ 2.0005    \\
\hline
 \end{tabular}
\label{table1}
\end{table} 

For what concerns the Kolmogorov constant $C$, it is defined by 
$
\ds E(\kappa) \approx C {\varepsilon^{2/3} } / {\kappa^{5/3}}
$,
and is here calculated as
\bea
C = \max_{\kappa \in (0, \infty)} \frac{E(\kappa) \kappa^{5/3}}{\varepsilon^{2/3}} 
\eea
In the table \ref{table1}, the Kolmogorov constant is shown in terms of the same Reynolds numbers.
The obtained values of $C$ increase with the Reynolds number and are in good agreement with the 
numerical and experimental values known from the various literature \cite{Kerr90}, \cite{Vincent91}, \cite{Yeung97}.

The spatial structure of $\Delta u_r$, expressed by Eq. (\ref{fluc4}), is also studied 
with the previous results. 
According to the various works \cite{Kolmogorov62}, \cite{Stolovitzky93}, \cite{She-Leveque94}, 
$\Delta u_r$ behaves quite similarly to a multifractal system, where $\Delta u_r$
obeys to a law of the kind 
$
\Delta u_r(r) \approx r^q
$
in which $q$ is a fluctuating exponent.
This implies that the statistical moments of $\Delta u_r(r)$ are expressed through 
different scaling exponents $\zeta(n)$ whose values depend on the moment order $n$, i.e.
\bea
\left\langle (\Delta u_r)^{n} \right\rangle  = A_n r^{\zeta(n)}
\label{fractal}
\eea
\suppressfloats
\begin{figure}[t]
	\centering
         \includegraphics[width=0.47\textwidth]{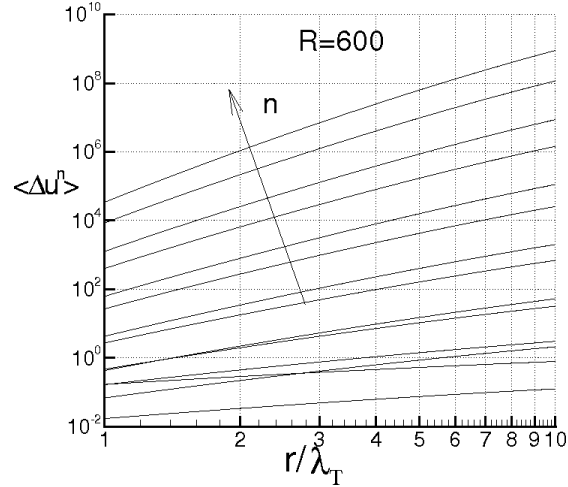}
\caption{Statistical moments of $\Delta u_r$ in terms of
the separation distance, for $R$=600.}
\label{figura_8}
\end{figure}
\suppressfloats
\begin{figure}[t]
\vspace{-0.mm}
	\centering
         \includegraphics[width=0.47\textwidth]{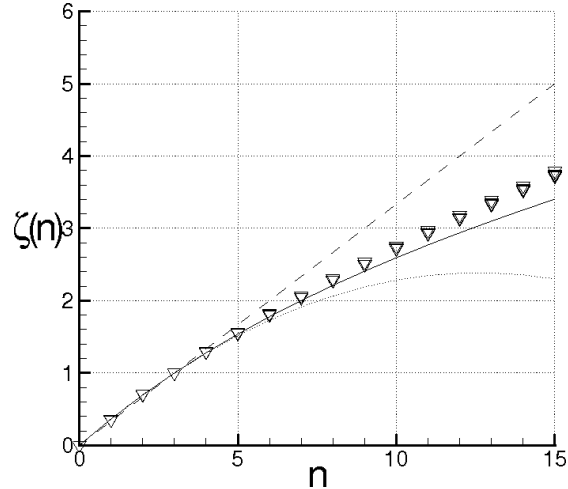}
\caption{Scaling exponents of $\Delta u_r$ for several $R$. Solid symbols are for the present data. Dashed line is for Kolmogorov K41 data \cite{Kolmogorov41}. Dotted line is for Kolmogorov K62 data \cite{Kolmogorov62}.  Continuous line is for She-Leveque data \cite{She-Leveque94}}
\label{figura_9}
\end{figure}
\suppressfloats
\begin{table}[t]
\centering
\caption{Scaling exponents of the longitudinal velocity difference for several Taylor-Scale Reynolds number.}
  \begin{tabular}{ccccccc} 
R           &  100  &  200 &  300  &  400  & 500  &  600 \\
\hline
\hline \\
$\zeta(1)$  &  0.35 & 0.35 &  0.35 &  0.35 & 0.35 &  0.35    \\ 
$\zeta(2)$  &  0.70 & 0.71 &  0.71 &  0.71 & 0.71 &  0.71    \\ 
$\zeta(3)$  &  1.00 & 1.00 &  1.00 &  1.00 & 1.00 &  1.00    \\
$\zeta(4)$  &  1.30 & 1.29 &  1.29 &  1.29 & 1.29 &  1.29    \\
$\zeta(5)$  &  1.56 & 1.55 &  1.55 &  1.55 & 1.55 &  1.55    \\
$\zeta(6)$  &  1.82 & 1.81 &  1.81 &  1.81 & 1.81 &  1.81    \\
$\zeta(7)$  &  2.06 & 2.05 &  2.05 &  2.04 & 2.04 &  2.05    \\
$\zeta(8)$  &  2.31 & 2.28 &  2.28 &  2.28 & 2.28 &  2.28    \\
$\zeta(9)$  &  2.53 & 2.50 &  2.50 &  2.50 & 2.50 &  2.51    \\
$\zeta(10)$ &  2.76 & 2.72 &  2.73 &  2.72 & 2.72 &  2.73    \\
$\zeta(11)$ &  2.97 & 2.93 &  2.93 &  2.93 & 2.93 &  2.94    \\
$\zeta(12)$ &  3.18 & 3.14 &  3.14 &  3.13 & 3.13 &  3.15    \\
$\zeta(13)$ &  3.39 & 3.33 &  3.34 &  3.33 & 3.33 &  3.35    \\
$\zeta(14)$ &  3.59 & 3.53 &  3.54 &  3.53 & 3.53 &  3.55    \\
$\zeta(15)$ &  3.79 & 3.73 &  3.73 &  3.72 & 3.73 &  3.75    \\
\hline
 \end{tabular}
\label{table2}
\end{table} 
\suppressfloats
\begin{figure}[b]
	\centering
         \includegraphics[width=0.50\textwidth]{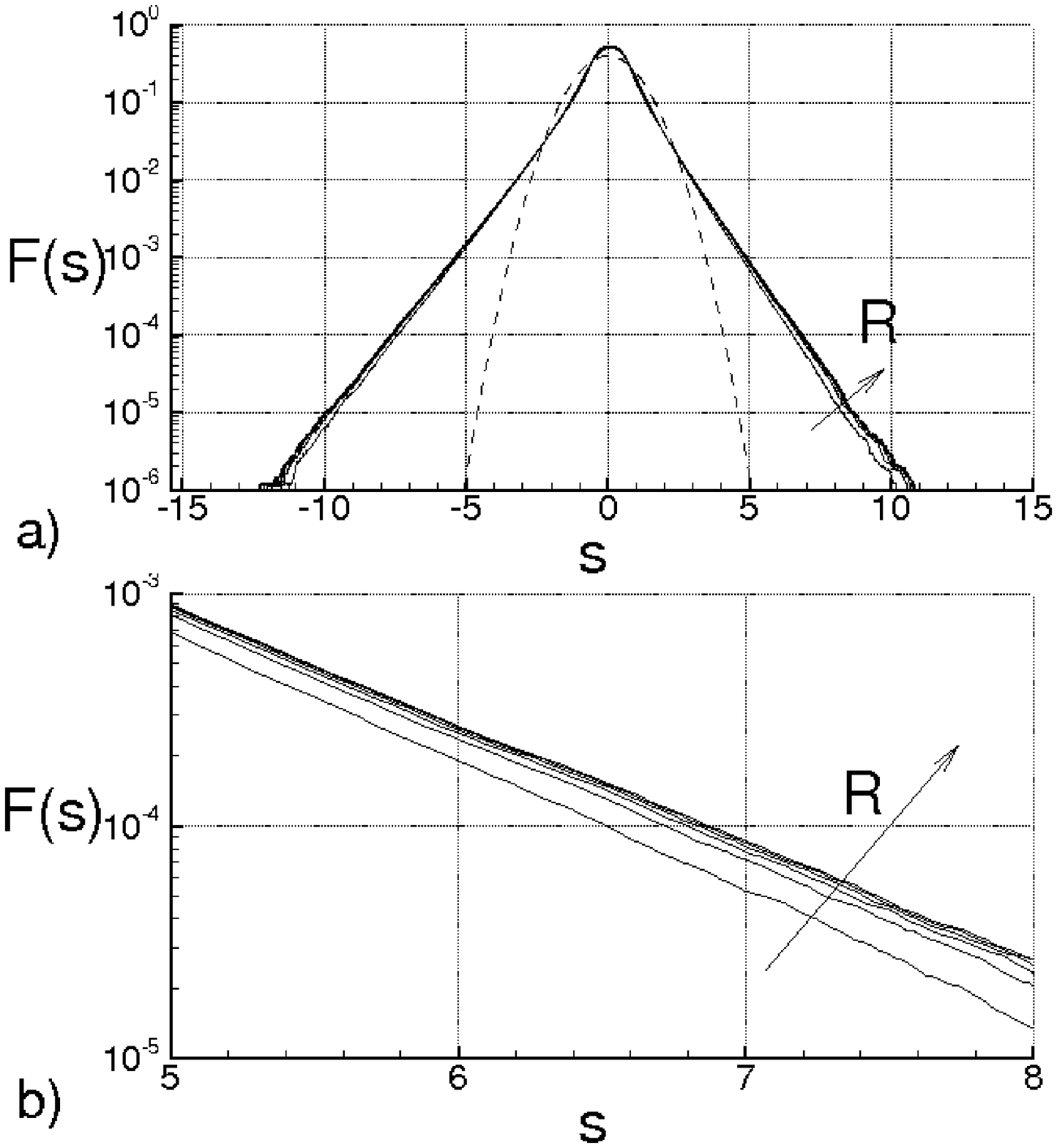}
\caption{Probability distribution functions of the longitudinal velocity derivative 
 for the different Taylor-Scale Reynolds numbers}
\label{figura_10}
\end{figure}
In order to calculate these exponents, the statistical moments of $\Delta u_r$ are first
calculated using Eqs.  (\ref{m1}) for several separation distances.
Figure \ref{figura_8} shows the evolution of the statistical moments of $\Delta u_r$ in terms of $\hat{r}$, in the case of $R = 600$.
The scaling exponents of Eq. (\ref{fractal}) are identified through a best fitting procedure,
in the intervals ($\hat{r}_1, \hat{r}_2$), where the endpoints $\hat{r}_1$ and $\hat{r}_2$ 
have to be determined.
The calculation of $\zeta(n)$ and $A_n$ is carried out through a minimum square method which,
for each moment order, is applied to the following optimization problem
\bea
\ds J_n(\zeta(n), A_n) \hspace{-1.mm} \equiv  
\int_{\hat{r}_1}^{\hat{r}_2} 
\ds ( \langle (\Delta u_r)^n \rangle - A_n r^{\zeta(n)} )^2 dr 
 = \mbox{min}, \   n = 1, 2, ...
\eea
where $(\langle (\Delta u_r)^n)\rangle$ are calculated with Eqs. (\ref{m1}),
$\hat{r}_1$ is assumed to be equal to $0.1$, whereas $\hat{r}_2$ is taken in such a way
that $\zeta(3)$ = 1 for all the Reynolds numbers.
The so obtained scaling exponents are shown in Table (\ref{table2}) in terms of the 
Taylor scale Reynolds number, whereas in Fig. \ref{figura_9} (solid symbols) 
these exponents are compared with those of the Kolmogorov theories K41 \cite{Kolmogorov41} (dashed line) and K62 \cite{Kolmogorov62} (dotted line), and with the scaling exponents calculated by She-Leveque \cite{She-Leveque94} (continuous curve).
Near the origin $\zeta(n) \simeq n/3$, and in general the values of $\zeta(n)$ are in good agreement with the She-Leveque results. In particular the scaling exponents here calculated
are lightly greater than those by She-Leveque for $n >$ 8.

According to the present analysis,
these peculiar laws $\zeta$ $ = \zeta(n)$ which make $\Delta u_r$ similar to a multifractal system, 
are the consequence of the combined effect of the quadratic terms into Eq. (\ref{fluc4}) 
and of the functions $K$ and $k$ calculated through  Eq. (\ref{vk-h2}).

The PDFs of $\partial u_r /\partial {\hat{r}}$ can be formally determined with 
Eqs. (\ref{frobenious_perron}) and (\ref{fluc4}).
Specifically, these PDF are calculated with several direct simulations, where the sequences of the variables $\xi$, $\eta$ and $\zeta$ are first determined by a gaussian random numbers generator.
The distribution function is then calculated through the statistical elaboration of the data obtained with Eq. (\ref{fluc4}).
The results are shown in Fig. \ref{figura_10}a and \ref{figura_10}b in terms of the dimensionless abscissa 
\bea
\ds s = \frac{\partial u_r /\partial {\hat{r}} } 
{ \langle \left( \partial u_r /\partial {\hat{r}} \right) ^2 \rangle^{1/2}  }
\nonumber
\eea
These distribution functions are normalized, in order that their standard 
deviations are equal to the unity. The figure represents the PDF for the several
$R$, and the dashed curve represents the gaussian distribution functions.
In particular, Fig. \ref{figura_10}b shows the enlarged region of Fig. \ref{figura_10}a,  
where  $5 < s < 8$. According to Eq. (\ref{fluc4}), although the skewness of
$\partial u_r /\partial {\hat{r}}$ does not depend on $R$, the tails of PDFs change with $R$ 
in such a way that the intermittency of $\partial u_r /\partial {\hat{r}}$ rises with
the Reynolds number. This increasing intermittency, caused by the quadratic terms appearing into
Eq. (\ref{fluc4}), is the result of the constancy of the skewness of  $\partial u_r /\partial {\hat{r}}$.

\section{\bf  Conclusions  \label{s9}}

The obtained self--similar solutions of the von K\'arm\'an-Howarth equation with
the proposed closure,  and the corresponding characteristics of the fully developed 
turbulence are shown to be in very good agreement with the various properties of the
turbulent flow from several points of view. 

In particular:

\begin{itemize}
\item 
The energy spectrum follows the Kolmogorov law in a range of wave-numbers
whose size increases with the Reynolds number, whereas for higher wave-numbers, it 
diminishes according to an exponential law.

\item
As the consequence of the skewness of velocity difference,
the Kolmogorov function exhibits a maximum and relatively small variations in proximity of 
$r = O(\lambda_T)$. 
This maximum value rises with the Reynolds number and seems to tend toward the limit
$4/5$, prescribed by the Kolmogorov theory.

\item 
The Kolmogorov constant moderately rises with the Reynolds number with an average
value around to 1.95 when $R$ varies from 100 to 600.

\item
The scaling exponents of the moments of velocity difference are calculated through
a best fitting procedure in an opportune range of the separation distance.
The values of these exponents are in good agreement with the results known 
from the literature.
\end{itemize}

These results represent a further test of the analysis presented in Ref. \cite{deDivitiis2009} 
which adequately describes many of the properties of the isotropic turbulence.

\section{\bf  Acknowledgments}

This work was partially supported by the Italian Ministry for the 
Universities and Scientific and Technological Research (MIUR).




\end{document}